\documentstyle[prl,aps,graphicx]{revtex}

\begin{document}

\title{Bose-Einstein Condensates in Time-Dependent Light Potentials:
Adiabatic and Nonadiabatic Behavior of Nonlinear Wave Equations}

\author{Y.\ B.\ Band$^{\,1}$ and Marek Trippenbach$^{\,1,2}$}
\address{
$^{\,1}$ Department of Chemistry, Ben-Gurion University of the Negev, 
84105 Beer-Sheva, Israel \\
$^{\,2}$ Institute of Experimental Physics, Optics Division,
Warsaw University, ul.~Ho\.{z}a 69, Warsaw 00-681, Poland
}

\maketitle

\begin{abstract}
The criteria for validity of adiabaticity for nonlinear wave equations
are considered within the context of atomic matter-waves tunneling
from macroscopically populated optical standing-wave traps loaded from
a Bose-Einstein condensate.  We show that even when the optical
standing wave is slowly turned on and the condensate behaves
adiabatically during this turn-on, once the tunneling-time between
wells in the optical lattice becomes longer than the nonlinear
time-scale, adiabaticity breaks down and a significant spatially
varying phase develops across the condensate wave function from well
to well.  This phase drastically affects the contrast of the fringe
pattern in Josephson-effect interference experiments, and the
condensate coherence properties in general.
\end{abstract}


{\it Introduction.} --- Recent matter-wave interference studies with
Bose-Einstein condensed atoms
\cite{Andrews,Hagley,Anderson,Orzel,Cata} have shown that one can
observe phase-dependent dynamics in dilute neutral-atom systems, in
complete analogy with those observed in Josephson-effect experiments
\cite{Josephson}.  In some of the experiments reported in
Refs.~\cite{Anderson,Orzel}, the external potential (applied by
optical standing waves to the initial ground-state Bose-Einstein
condensate (BEC)) is slowly turned on.  One might assume, based upon
the Adiabatic Theorem of Quantum Mechanics \cite{Messiah,CohenTan},
that the system should remain in an eigenstate, even though the nature
of the eigenstate evolves in time.  Moreover, adiabaticity criteria
for nonlinear wave equations have been studied in connection with
soliton dynamics in nearly integrable systems, and one expects on the
basis of these studies that adiabaticity should be maintained with
slow enough variation of the parameters of the system \cite{KM}.  A
spatially varying phase of the condensate wave function is a
manifestation of non-adiabatic dynamics of a BEC, in the sense that
the adiabatic eigenstate (the ground nonlinear eigenstate of the
Gross-Pitaevskii equation (GPE) calculated using the potential at any
instant of time) can be taken to be real (it does not have a spatially
depenendent phase).  Hence one concludes that, a spatially varying
phase should not develop across the condensate in
Refs.~\cite{Anderson,Orzel} when the optical potential is turned on
very slowly.  Here we present calculations corresponding to conditions
similar to those reported in \cite{Anderson,Orzel} showing that the
BEC remains adiabatic as the light potential is turned on very slowly,
and the phase is constant across the condensate.  The dynamics are
indeed adiabatic despite the fact that nonlinear dynamics precludes
the possibility of a superposition principle, which is used heavily in
deriving the Adiabatic Theorem.  We further show that, as the light
potential is slowly increased in strength to the point where the wave
packets in the individual optically-induced wells become strongly
separated, and the time-scale for tunneling becomes long compared to
the nonlinear time-scale, adiabaticity is destroyed and a large
spatially varying phase develops across the wave function.  This large
inhomogeneous phase, obtained within a mean-field approximation, is
due to the tunneling dynamics (induced by the kinetic energy of the
BEC in the lattice potential) not being able to implement the
equilibration of the phase across the BEC on the nonlinear time-scale.

The process of dynamic splitting of a condensate by an external
time-varying potential was studied using two mode models
{\cite{Menotti} which show that the mean-field approximation based
upon the GPE will not give a good description of the splitting process
due to the slow rise of a potential barrier which cuts off tunneling
processes.  Our results show that, even within a mean-field
approximation, a sudden inhomogeneous phase build-up begins to develop
when the tunneling time becomes comparable to or larger than the
nonlinear time-scale, and the process of optical lattice turn-on
ceases to be adiabatic when this happens.  A transition occurs when
these two time-scales become comparable, already within a mean-field
GPE approach.  The spatially varying phase adversely affects the
fringe contrast in interference experiments performed on the BEC
wavepackets.

Adiabaticity in nonlinear systems can be studied in various regimes. 
Denoting $\tau_{AD}$ as the quantum-mechanical linear adiabatic
time-scale determined in terms of the difference of the (linear)
energy eigenvalues \cite{Messiah} and $\tau_{NL}$ the nonlinear
time-scale \cite{Tripp_2000} (see below), the simplest regime is one
in which the the duration of the dynamical process being studied, $T$,
satisfies the condition, $\tau_{AD}\ll T\ll \tau_{NL}$.  In this case,
adiabaticity is insured by the Adiabatic Theorem due to the first
inequality, and nonlinearity cannot play a significant role in the
dynamics due to the second inequality.  Hence, the dynamics must be
adiabatic.  The regime in which the experiments of
Refs.~\cite{Anderson,Orzel} are carried out satisfy the condition
$\tau_{AD}, \tau_{NL} \ll T$, and the nonlinearity does play an
important role in the dynamics.  Nevertheless, as described below, the
dynamics are indeed adiabatic, until the strength of the optical
lattice is so large that the wave packets in the individual wells
become almost completely separated.

We consider Bose-Einstein condensed $^{87}$Rb atoms in the $|F=2,
M_F=2\rangle$ hyperfine state confined in an array of optical traps in
a gravitational field.  The atoms are trapped at the antinodes of a
vertically oriented red-detuned optical standing wave, which are
separated by $\Delta z = \lambda/2$, where $\lambda = 840$ nm is the
wavelength of light used to confine the atoms.  The depth of the
optical potential is proportional to the intensity of the light; the
intensity is initially zero and linearly increases with time.  The
initial BEC is cigar shaped with 10$^4$ atoms in a static harmonic
trap potential $V_{ho}({\bf r}) = \frac{m\omega_z}{2} z^2 +
\frac{m\omega_{x,y}}{2} (x^2+y^2)$ with frequencies $\omega_z = 2\pi
\, \times \, 19$ Hz, and $\bar{\omega} = 2\pi \, \times \, 33$ Hz
($\bar{\omega} = (\omega_x\omega_y\omega_z)^{1/3}$ with
$\omega_x=\omega_y \equiv \omega_{xy}$) \cite{Orzel}.  The
light-potential experienced by the atoms in the BEC is given by
$V_L(z,t) = V_0(t) (1+\cos(2k_Lz))$ where the well-depth, $V_0(t)=125
\, t\, [E_R]$, varies linearly with time.  The recoil energy $E_R =
\frac{\hbar^2 k_L^2}{2m}$ is the kinetic energy gained by an atom
absorbing a photon from the optical lattice, where $m$ is the atomic
mass, and the photon wavevector is $k_L = 2\pi/\lambda$.  The rate of
increase of the light-potential, $(125 \ [E_R/s])$, is sufficiently
slow, as shown below, that the dynamics of the BEC is adiabatic for
much of the turn-on.  After some time, the harmonic potential and the
light-potential are switched-off (dropped) releasing the atoms to fall
under the influence of gravity.  In the experiments reported in
Ref.~\cite{Orzel}, the atoms are held in the optical lattice for a
short time (2.5 ms) after switching off the harmonic potential,
allowing the gravitational potential difference between wells to
affect the phase difference between wells, and absorption images are
taken 8 ms after the optical potential is turned off and the atoms
begin to free-fall.

{\it Theoretical Formulation} --- The mean-field dynamics can be
determined in terms of the time-dependent GPE, $i\hbar
\frac{\partial\psi({\bf r},t)}{\partial t} = (p^2/2m + V({\bf r},t) +
N_0 U_0|\psi|^2) \psi$, where $V({\bf r},t) = V_{ho}({\bf r}) - mgz +
V_L(z,t)$, $U_{0} = \frac{4\pi a_{0}\hbar^{2}}{m}$ is the atom-atom
interaction strength that is proportional to the $s$-wave scattering
length $a_{0}$, and $N_0$ is the total number of atoms.  We solve the
time-dependent GPE using a split-operator fast Fourier transform
method to propagate an initial state of the BEC in time in the
presence of the harmonic potential, gravity and the time dependent
optical lattice; the initial state is determined by propagating in
imaginary time with vanishing optical lattice potential
\cite{Tripp_2000}.  Due to the large number of grid points necessary
in the lattice direction ($z$), and the large number of time steps
necessary to propagate the GPE to completion of the dynamics, we found
it necessary to convert the 3D GPE into an effective 1D GPE with
similar dynamics.  This is carried out using the following procedure. 
The 1D GPE is written in terms of characteristic time-scales $t_{DF}$
for diffraction and $t_{NL}$ for the nonlinear interaction, in the
following manner \cite{Tripp_2000,Tripp98}:
\begin{equation}
\frac{\partial\psi}{\partial t} = i\left[ \frac{r_{TF}^2}{t_{DF}} \,
\frac{\partial^2 }{\partial z^2} - V(z,t)/\hbar -
\frac{1}{t_{NL}} \,\frac{|\psi|^2}{|\psi_{m}|^{2}} \right] \psi \ .
\label{GP_reduced}
\end{equation}
Here, $r_{TF} = \sqrt{2\mu/(m\bar{\omega}^2)}$ is the Thomas-Fermi
radius, $\mu$ is the initial chemical potential, $\mu = \frac{1}{2}
\left( \frac{15 U_0 N}{4\pi}\right)^{2/5} (m\bar{\omega}^2)^{3/5}$,
$t_{DF}$ is the diffraction time $t_{DF} = 2m r_{TF}^2/\hbar$ and
$|\psi_{m}|$ is the maximum initial magnitude of the wave function
defined in terms of the initial nonlinear time by $t_{NL} =
(\mu/\hbar)^{-1} = (NU_0 |\psi_{m}|^2 /\hbar)^{-1}$ \cite{Tripp_2000}. 
We take all lengths in the GPE (\ref{GP_reduced}) in units of the
Thomas-Fermi radius along the $z$ axis, and we call this length unit
$r_z \equiv r_{TF,z}$.  The key to obtaining physically relevant
dynamics using the 1D GPE, is to use (1) $\omega_{1D} = (\frac{2\mu}{m
r_z})^{1/2}$, and (2) $N_{1D} = N_0 (\bar{\omega}/\omega_{1D})^3$.
These two equations insure that the Thomas-Fermi radius and the
nonlinear time (or the chemical potential) remain as in 3D world.  We
checked to confirm that this procedure gives the same wave function
as the 1D projection of the 3D GPE solution for our studies using
shorter propagation times.

{\it Numerical Results} --- Starting from the BEC without any optical
lattice present, we begin to increase the optical potential with the
linear ramp $V_0(t)$ mentioned earlier.  Fig.~\ref{fig1} shows the
magnitude and phase of the Gross-Pitaevskii wave function, $\psi(z,t)
= |\psi(z,t)| \exp(i\theta(z,t))$ as a function of position in the
optical lattice, $z/r_z$, when the well-depth of the optical lattice
is 10 and 15 $E_R$.  We have taken $z=0$ to be at the minimum of the
harmonic potential plus the gravitational potential, i.e., we have
shifted the definition of the center of the trap to the true center of
the combined harmonic plus gravitational potential.  About 17 wells of
the optical potential are populated during the course of the dynamics. 
The magnitude $|\psi(z,t)|$ remains normalized throughout the
propagation; if we were to average out the oscillations in
$|\psi(z,t)|$ we would obtain roughly the initial BEC $|\psi(z,0)|$. 
The phase $\theta(z,t)$ is nearly spatially independent, and the
magnitudes $|\psi(z,t)|$ in Fig.~\ref{fig1} are very nearly equal to
the eigenstates of the GPE with the optical potential present at these
times.  The slight variation of the phases with position $z$ indicate
the small degree of the nonadiabaticity that results during the
dynamics.  The phases $\theta(z,t)$ vary as a function of time by a
spatially independent constant due to the dynamics but this
($z$-independent) constant phase is not physically significant in the
experiments of Refs.~\cite{Anderson,Orzel}.  The phase $\theta(z,t)$
at the time when the depth of the optical lattice is 10 $E_R$ has been
shifted up in Fig.~\ref{fig1} by unity so that it could be plotted
conveniently.

As we increase the well-depth of the optical lattice further, a regime
is reached in which the wave packets localized in the various wells
become almost completely separated.  When this occurs, the spatially
varying phase of the wave function begins to grow significantly, and
the spatially dependent variations increase as the well-depth of the
optical lattice increases.  Fig.~\ref{fig2} shows the magnitude
$|\psi(z,t)|$ and phase $\theta(z,t)$ of the GP wave function as a
function of position in the optical lattice, $z/r_z$, when the
well-depth of the optical lattice is 25 and 40 $E_R$.  The magnitude
of the wave function is almost totally within the region $z/r_z \in
[-1.3,1.3]$.  The wave packets in the optical wells are almost fully
separated.  The spatially dependent variations in the phase are less
pronounced in the center of the trap where the variation in the peak
densities of the wave packets from well to well are smaller; by 40
$E_R$ the variation in phase across the condensate is large compared
with 1 radian.  The jump in the phase for 40 $E_R$ at $z/r_z \approx
\pm 1.1$ is artificial and due to the continuation of the inverse
trigonometric function used to calculate $\theta(z,t)$ from
$\psi(z,t)$.  The spatial phase variation over the condensate for 50
$E_R$ is much greater than that for 40 $E_R$; it varies over more 
than $5 \pi$.

Fig.~\ref{fig3} shows the magnitude of the Fourier transform of the
Gross-Pitaevskii wave function $\psi(k)$ versus $k$ when the
well-depth of the optical lattice is 25, 40 and 50 $E_R$.  The wave
function has amplitude around $k = 0$, $\pm 2 k_L$, although
components around $k = \pm 4 k_L$ are also clearly visible.  The width
of the wavepackets increase with increasing well-depth, particularly
as the spatially varying phase across the condensate becomes
significant .  By 50 $E_R$, the width of Fourier components of the wave
packets are a good fraction of $2 k_L$.

Adiabaticity is maintained throughout the course of the dynamics until
the wave packets become well separated and the tunneling time becomes
comparable to or larger than the nonlinear time-scale.  Adiabaticity
of soliton solutions of nonlinear wave equations have been extensively
studied for slowly varying external conditions (for a review, see
Ref.~\cite{KM}).  It has been shown that if the time-scale $T$ of the
variation of an external parameter is slow compared the instantaneous
nonlinear eigenvalue $\omega_0(t)$ of the nonlinear (time-independent)
equation at time $t$, $\omega_{0}(t) T/(2\pi )\gg 1$, the dynamics can
be adiabatic.  Here we see that an additional condition is required;
adiabaticity breaks down when the wave packets become well separated
and the tunneling time becomes longer than the nonlinear time.

It is easy to make a rough estimate of the tunneling probability
from well to well using the semiclassical approximation
\cite{CohenTan},
\begin{equation}
P(t) = \exp \left(-\int_0^{\pi/(2k_L)} k(z,t) dz \right) \approx \exp
\left(- \sqrt{\frac{2m}{\hbar^2}} \int_0^{\pi/(2k_L)} (V_L(z,t))^{1/2}
dz \right) = \exp \left(-2\sqrt{4mV_0(t)}/(\hbar k_L) \right) \ .
\end{equation}
The tunneling rate is then given by $R(t) = \omega_v(t) \times P(t)$,
where the vibrational frequency $\omega_v(t)$ can be approximated by
$\omega_v(t) \approx \sqrt{4k_L^2 V_0(t)/m}$, since the expansion of
the optical potential about a minimum in the potential yields
$V_L(z,t) \approx (m/2)(\omega_v(t))^2 z^2 = (m/2)(4 k_L^2 V_0(t)/m)
z^2$.  Hence, the time-dependent tunneling rate is $R(t) = \sqrt{8
(V_0(t)/E_R)} \, \exp \left(- \sqrt{8(V_0(t)/E_R)} \right) \,
[E_R/\hbar]$.  The tunneling rate $R(t)$ can be compared with the
time-dependent inverse nonlinear time-scale, $(t_{NL}(t))^{-1} = NU_0
|\psi_{m}(t)|^2 /\hbar = \mu(t)/\hbar$ and the time-dependent
diffraction time-scale, $(t_{DF}(t))^{-1} = \frac{\hbar
(\omega_v(t))^2}{4\mu(t)}$.  Once the tunneling time, $R^{-1}$,
becomes long compared to $t_{NL}$ (and $t_{DF}$, which is typically
the smallest of these time-scales), a spatially independent phase can
not be maintained across the condensate by the action of the kinetic
energy operator, and a spatially dependent phase develops.  In our
calculations, this happens beyond a lattice well-depth of about 35
$E_R$.

If we take our calculated wave packets (shown in Figs.~\ref{fig1} and
~\ref{fig2}) and propagate for an additional 2.5 ms upon switching off
the magnetic field, but leaving the optical lattice potential in
place, as described in Ref.~\cite{Orzel}, our wave packets hardly
change.  However, upon switching off the optical potential and
propagating the wave packets for 8 ms, the wave packets spread by
diffusion very significantly as they fall in gravity.  The resulting
wave packets are considerably different upon using the 25 and 50 $E_R$
results.  The additional Fourier components of the wave packet in the
50 $E_R$ case significantly wash out interference patterns in the
density profile for this case, consistent with the measurement in
\cite{Orzel}.  Note that we are not asserting that squeezing is absent
in the experiments of Ref.~\cite{Orzel}; rather, that an improved
model of the above-the-mean field effects that accounts for the
spatially varying phase is necessary to quantitatively compare with
the experiments.

{\it Conclusions} --- We have seen that the spatially varying phase of
the solution to the Gross-Pitaevskii equation becomes large when the
tunneling time between wells in the optical lattice becomes comparable
to or larger than the nonlinear time-scale.  Adiabaticity then breaks
down and the instantaneous nonlinear eigenvectors to the
time-independent Gross-Pitaevskii equation do not have the character
of the dynamical solution.  This nonadiabaticity takes place even when
the variation of the optical potential is sufficiently slow that
adiabaticity is otherwise assured.  It remains to determine the
above-mean-field corrections to this picture {\em using an approach
that incorporates an accurate form for the spatially varying complex
order parameter}.

Finally, we note that these results have implications regarding the
loading of an optical lattice with atoms for quantum computing, using a
Bose-Einstein condensate (BEC) source and laser beams that are slowly
turned on ~\cite{Brennen_Jaksch}.  From our studies we conclude that
the laser fields must be controlled so that the density of the atoms
is sufficiently low (the nonlinear term must be negligible) {\em
before} tunneling between the wells is cut off by the full depth of
the optical potential.  Otherwise, phase variations from well to well
will deleteriously affect the resulting optical lattice state.

\bigskip

Useful conversations with Boris Malomed and Paul Julienne are
gratefully acknowledged.  This work was supported in part by grants
from the U.S.-Israel Binational Science Foundation (grant No. 
98-421), Jerusalem, Israel, the Israel Science Foundation (grant No. 
212/01), the Israel MOD Research and Technology Unit and the Polish
KBN 2/P03/B07819

\begin{figure}[!htb]
\centerline{\includegraphics[width=3in,angle=270,keepaspectratio]{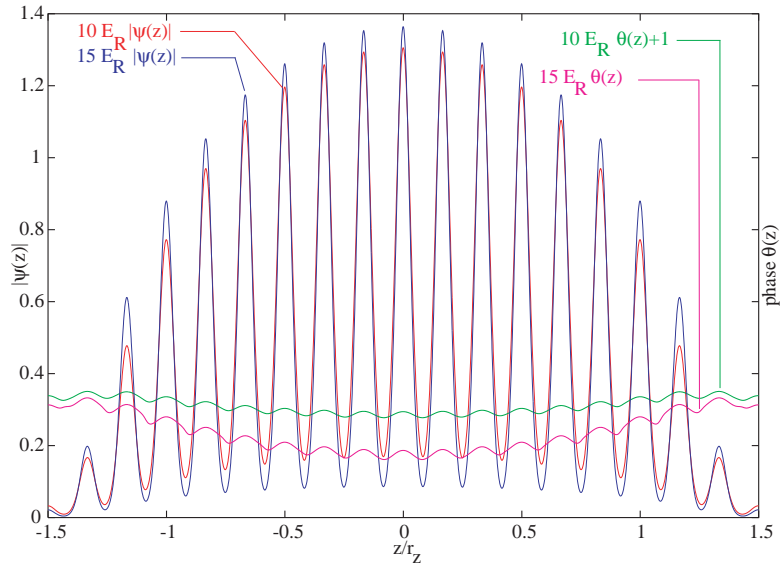}}
\caption {Magnitude and phase of the Gross-Pitaevskii wave function as
a function of position in the optical lattice.  The wave function is
shown when the well-depth of the optical lattice are 10 and 15 $E_R$.}
\label{fig1}
\end{figure}

\begin{figure}[!htb]
\centerline{\includegraphics[width=3in,angle=270,keepaspectratio]{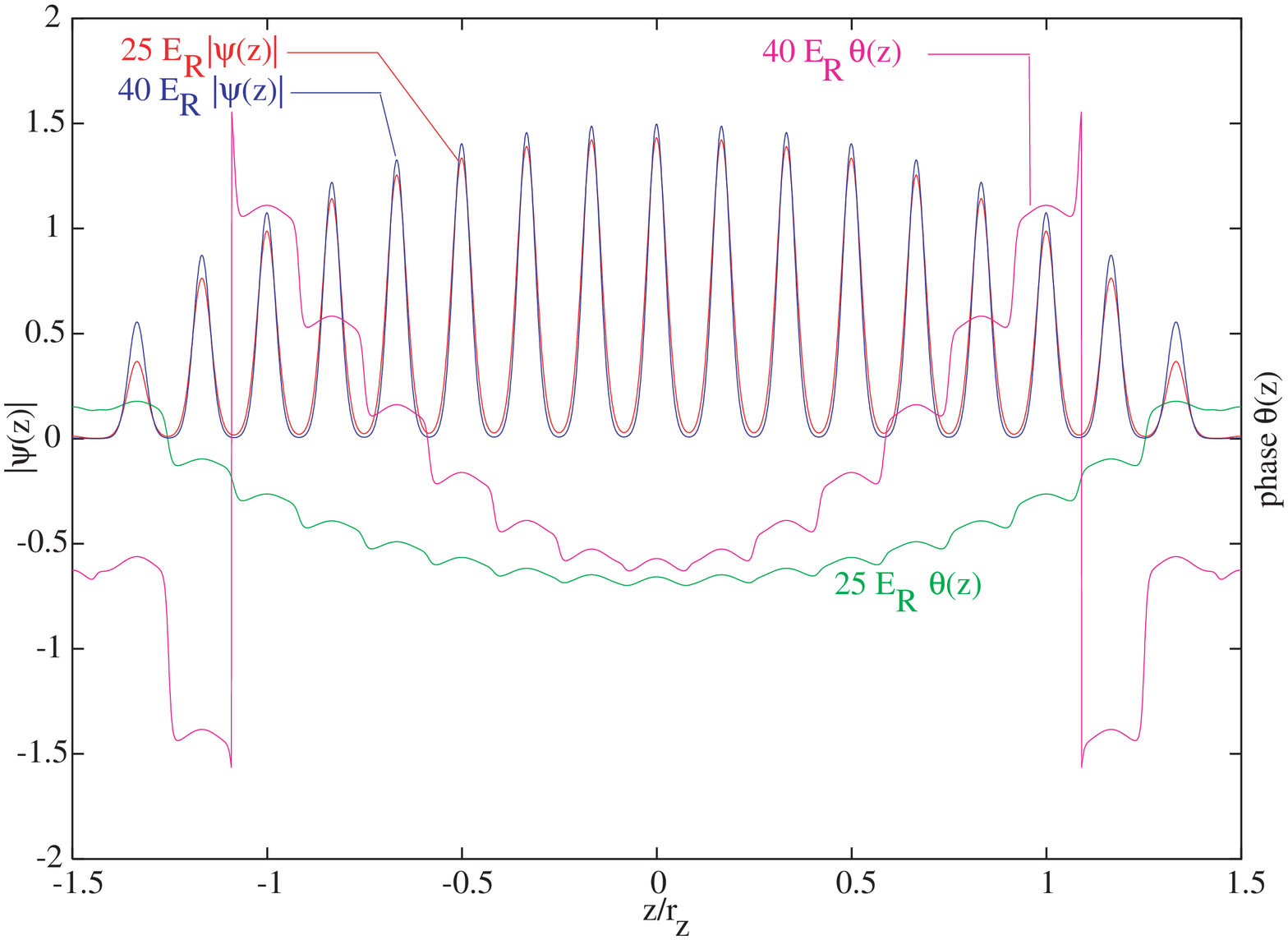}}
\caption {Magnitude and phase of the Gross-Pitaevskii wave function as
a function of position in the optical lattice.  The wave function is
shown when the well-depth of the optical lattice are 25 and 40 $E_R$.}
\label{fig2}
\end{figure}

\begin{figure}[!htb]
\centerline{\includegraphics[width=3in,angle=270,keepaspectratio]{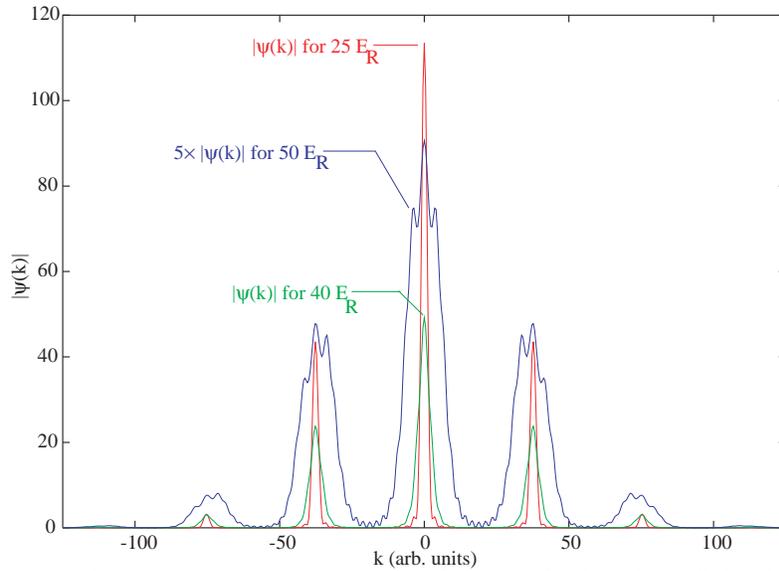}}
\caption {Gross-Pitaevskii wave function in momentum space, $\psi(k)$
versus $k$ when the well-depth of the optical lattice are 25, 40 and
50 $E_R$.}
\label{fig3}
\end{figure}


\begin{references}

\bibitem{Andrews} 
M. R. Andrews {\it et al.}, Science 275, {\bf 637} (1997).

\bibitem{Hagley}
E. W. Hagley, {\it et al.}, Phys. Rev. Lett. {\bf 83}, 3112 (1999);
M. Trippenbach, {\it et al.}, J. Phys. B {\bf 33}, 47-54 (2000).

\bibitem{Anderson} B. P. Anderson and M. A. Kasevich, Science {\bf 282}, 
1686 (1998).

\bibitem{Orzel} C. Orzel {\it et al.}, Science {\bf 291}, 2386 (2001).

\bibitem{Cata}
F. S. Cataliotti {\it et al.}, Science {\bf 293}, 843 (2001).

\bibitem{Josephson} A. Barone and G. Paterno, {\it Physics and
Applications of the Josephson Effect} (Wiley, New York, 1982).

\bibitem{Messiah} A. Messiah, {\it Quantum Mechanics}, Vol. II, Chp. 
17, (N. Holland, Amsterdam, 1975).

\bibitem{CohenTan} C. Cohen-Tannoudji, B. Diu and F. Lalo\"{e}, 
{\it Quantum Mechanics}, (John Wiley, NY, 1977).

\bibitem{KM} Y.S. Kivshar and B.A. Malomed, Rev. Mod. Phys. {\bf 61}, 763
(1989).

\bibitem{Menotti} C. Menotti, J.R. Anglin, J.I. Cirac and P. Zoller,
Phys. Rev. A{\bf 63}, 023601, 2001; A. J. Leggett and F. Sols, Phys. Rev. 
Lett. {\bf 81}, 1344 (1998); R.W. Spekkens and J.E. Sipe, Phys. Rev. {\bf 
A 59}, 3868 (1999); J. Javanainen and M.Yu. Ivanov, Phys. Rev. {\bf A 60}, 
2351 (1999); A. Vardi and J.R. Anglin, Phys. Rev. Lett. {\bf 86}, 
568 (2001).

\bibitem{Tripp_2000} M. Trippenbach, Y. B. Band, and P. S. Julienne, 
Phys. Rev. {\bf A62}, 023608 (2000).

\bibitem{Tripp98} M.\ Trippenbach, Y.\ B.\ Band, and P.\ S.\ Julienne,
Optics Express {\bf 3}, 530 (1998).

\bibitem{Brennen_Jaksch} G.K. Brennen, C.M. Caves, P.S. Jessen, and I.H. 
Deutsch, Phys. Rev. Lett. {\bf 82}, 1060 (1999); D. Jaksch, {\it et al.}, 
Phys. Rev. Lett. {\bf 82}, 1975 (1999).

\end{references}
\end{document}